# DESIGN AND IMPLEMENTATION OF THE CONNECTIONLESS NETWORK PROTOCOL (CLNP) AS LOADABLE KERNEL MODULES IN LINUX KERNEL 2.6

Bunga Sugiarto [1]), Danny Laidi [1]), Arra'di Nur Rizal [1]), Maulahikmah Galinium [1]), Pradana Atmadiputra [1]), Melvin Rubianto [1]), Husni Fahmi [2]), Tri Sampurno [2]), Marsudi Kisworo [3])

*Abstract*

*In this paper, we present an implementation of CLNP ground-to-ground packet processing for ATN in Linux kernel version 2.6. We present the big picture of CLNP packet processing, the details of input, routing, and output processing functions, and the implementation of each function based on ISO 8473-1. The functions implemented in this work are PDU header decomposition, header format analysis, header error detection, error reporting, reassembly, source routing, congestion notification, forwarding, composition, segmentation, and transmit to device functions. Each function is initially implemented and tested as a separated loadable kernel module. These modules are successfully loaded into Linux kernel 2.6.*

**Keywords:** ATN, CLNP, Linux Kernel

## 1. Introduction

Airplane usage as one of the means of transportation has greatly increased. Nevertheless, the support for air navigation and infrastructure of the aviation system is reaching its limits and soon will not be able to cope with the growing demand of air traffic. With this concern, ICAO, on 11[th] May 1998, at the Official Opening of the Worldwide CNS/ATM Systems Implementation Conference in Rio de Janeiro, instructs all participants of the Chicago conference which was held in 1944, including Indonesia, to deploy a sole CNS/ATM by 2015 [3]. As stated in Technology Roadmap of The Agency for the Assesment and Application of Technology (BPPT), BPPT supports Directorate General of Civil Aviation, PT. Angkasa Pura I, and PT. Angkasa Pura II to implement CNS/ATM in Indonesia [6]. Therefore, this research is an initial part of BPPT's research in order to provide safety in Indonesia air traffic.

A necessary part of CNS/ATM is the ATN (Aeronautical Telecommunication Network) which is an international communication infrastructure that manages digital data transfer between aircraft and civil air traffic control facilities.

This paper gives detail design and implementation of CLNP (input, routing, and output processes) inside Linux kernel version 2.6. Currently, the only available free source for CLNP is implemented using NetBSD [5]. Linux kernel version 2.6 is chosen because this is the latest stable version of Linux kernel when this paper is written.

---

[1] Swiss German University (SGU), Campus German Centre, Bumi Serpong Damai – 15321, Island of Java, Indonesia, {bunga.sugiarto | danny.laidi | arra.rizal | maulahikmah.galinium | pradana.priyono}@student.sgu.ac.id
[2] The Agency for the Assesment and Application of Technology (BPPT), Jl. MH. Thamrin no 8, Jakarta 13340, {fahmi | sampurno }@inn.bppt.go.id
[3] Swiss German University (SGU), Campus German Centre, Bumi Serpong Damai – 15321, Island of Java, Indonesia, {marsudi.kisworo}@sgu.ac.id

Functions in input, routing, and output processes are implemented separately as a loadable kernel module and tested by loading each of them manually into the Linux kernel. No performance measurement is done yet toward this research.

## 2. Related Work

We develop this network protocol based on NetBSD CVS Repositories [5] and IPv4 implementation in Linux kernel [2].

The NetBSD CVS Repositories provides the source code for the implementation of the network layer and transport layer of the ATN [5]. The source code files from the repositories that are closely related to our work are mbuf.h, clnp.h, clnp_input.c, clnp_er.c, clnp_frag.c, iso_chksum.c, clnp_output.c. This source code implements the CLNP functions using C programming language for the Net BSD kernel. While our work implement these functions using C programming language for the Linux kernel.

Rio et al explains the structure and organization of the networking code of Linux kernel 2.4.20 [2]. They explain the main data structure, the sub-IP layer, the IP layer, and two transport layers: TCP and UDP. This paper contributes our understanding of how the CLNP module can be integrated inside the Linux kernel, just like the IP module integrated in the same Linux kernel.

## 3. CLNP Design

This paper illustrates the design in the following order. Section 4.1 describes the overall CLNP big picture. Section 4.2, 4.3, and 4.4 explains CLNP input, routing, and output process respectively. Section 4.5 illustrates the integration of CLNP inside Linux kernel version 2.6. To comply with the international standard, the design for this research comes from ISO/IEC 8473-1 [4].

### 3.1. CLNP Big Picture

Figure 1 illustrates the CLNP big picture. To focuses on reliability and accurateness, the CLNP implementation is divided into three parts: input, routing, and output. Each part will be described in the following sections. We do not implement security, quality of service maintenance, echo request, and echo response functions.

### 3.2. CLNP Input Process

To give a clear overview, please refer to the CLNP input design illustrated in CLNP big picture (figure 1, boxed with an "input" label).

When a PDU is received, the header part and payload part of the PDU are decomposed. Then, the header part is checked and analyzed. An error in PDU can be detected using checksum calculation. Next, if segmentation part exists, the PDU will be reassembled first. After that, the destination address is checked. If the packet belongs to others, it will be forwarded. In contrast, the packet will be delivered to the transport layer.

PDU header error detection function detects an error inside the PDU header by checking the checksum parameters. If an error is detected, the PDU will be discarded and an error report will be generated and sent back to the receiver only if the Error Report (ER) flag inside PDU header is set to one. Otherwise, no error report will be generated. Error report PDU is similar to the normal PDU, except an error report PDU contains a reason for discard field. The comparison between a normal PDU and an error report PDU structure can be seen in ISO 8473-1 page 29 and 31 [3].

In case of PDU segmentation, reassembly function reconstructs the initial PDU from the segmented PDUs. Fragments are identified coming from the same initial PDU by their data unit identifier, destination address, and source address. The order of the packet is determined by the offset field. The reassembly function implemented in this research is proven capable of handling overlaps and segment duplication.

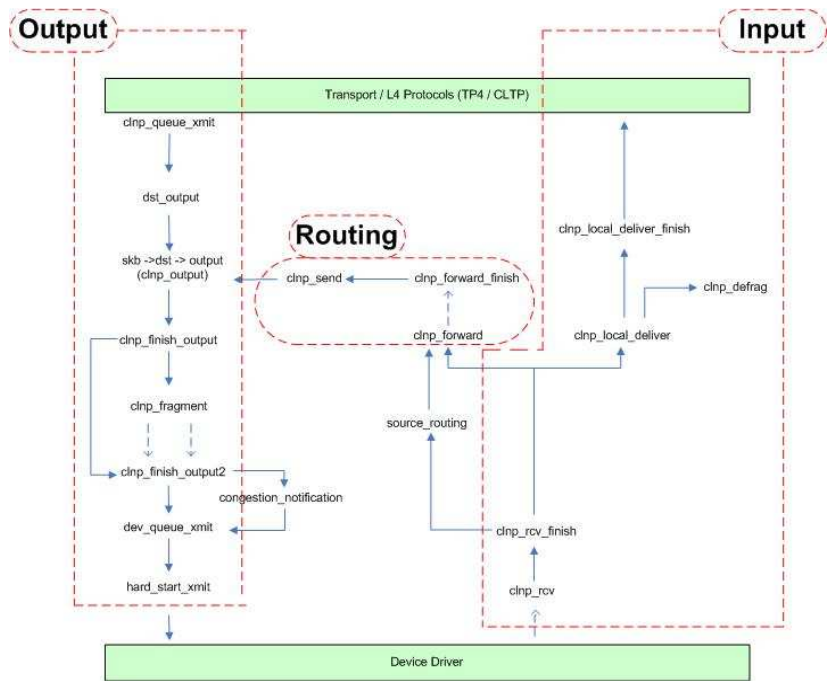

**Figure 1. CLNP Big Picture**

All of these functions are implemented successfully in this research. Decomposition and header format analysis functions are implemented in *clnp_rcv(), clnp_rcv_finish(), clnp_local_deliver(), and clnp_local_deliver_finish()* methods. Header error detection, discard PDU, and error report generation functions are implemented respectively in *clnp_check_csum(), discard_pdu(),* and *clnp_emit_er()* methods. Reassembly function is implemented by *clnp_find(), clnp_new_pkt(), clnp_insert_frag(),* and *clnp_comp_pdu()* methods.

### 3.3. CLNP Routing Process

A PDU can be destined from one End System (ES) to another End System (ES) through a series of Intermediate Systems (ISs). In CLNP, the route may or may not be determined initially by the originating network entity. If the route is to be determined by the originating network entity, the source routing parameter code inside the CLNP header is set to a value, and source routing function is called. At each ES or IS, CLNP needs to find the next hop to which the PDU should be forwarded to reach the destination ES. Therefore, all ESs and ISs must recognize each others' existence and reachability. This routing information is saved inside a routing table and accomplished by these routing protocols: ES-IS Routing Protocol, IS-IS Routing Protocol, and Inter Domain Routing Protocol. These three protocols deserve another paper. In this paper we implement forward PDU and source routing functions.

Forward PDU function forwards PDU to the appropriate host if the destination address is different from the local interface [1]. The forward PDU function receives the CLNP packet as the input from CLNP input process, processes the PDU, and sends it to CLNP output. According to Benvenutti [1], there are several processes which must be accomplished by the forward function. First, CLNP header must be analyzed. Then, a new buffer is allocated and the packet is copied into the new buffer. After that, the time to live (TTL) in the clnp header will be decreased. Since there is a change in the TTL, the checksum must be adjusted. Forwarding function also handles fragmentation when the PDU length is more than MTU (Maximum Transmission Unit). These whole processes are implemented in clnp_forward method. The last process is sending the packet to the outgoing device. This process is implemented in clnp_send method that will call clnp_output method.

The source routing is not a mandatory function. It is called only if it is selected by the originator of PDU. The source routing function allows a network entity to specify the path that a generated PDU shall take [2]. Then it will examine the route in the parameter value that defined by the network entity where the PDU constructed. The source route function will give the forward PDU function the data packet and the address of the next network entity where the packet needs to be transmitted. Source route analysis divided into complete source route analysis and partial source route analysis.

The source route analysis function examines the route in the parameter value that is defined by the network entity where the PDU is constructed. The Source Route Analysis will check whether the next network entity is reachable or not. If the next network entity is not reachable then the data packet will be discarded.

The function that checks the reachability of the next network entity is lookup function. The lookup function checks the routing table whether the entry (next network entity) is reachable or not. The lookup function that has been implemented so far is using sequential search. More efficient search should be developed for further research.

The source route analysis function is implemented in *source_route_analysis()* which will call *complete_source_route_analysis()* and *partial_source_route_analysis()* methods. The lookup function is implemented in *look_up()* method. This function will be called inside *complete_source_route_analysis()* and *partial_source_route_analysis()* methods.

### 3.4. CLNP Output Process

The output part of the CLNP processes packets that will be sent out of a system. The CLNP output process consists of PDU composition, PDU segmentation, and PDU transmission to device functions. The overview of the output part of CLNP is also given in Figure 1 (boxed with an "output" label).

PDU composition occurs when a packet arrives from the transport layer. First, the route of the packet to the destination address is determined by the routing mechanism. The routing mechanism in CLNP is described in the section 4.3 above. The routing data is then included in the options part of the CLNP header. After that, the PDU composition is performed by attaching CLNP header to the packet.

Whenever the size of a PDU is greater than the maximum transmission unit of the carrying network, it will be divided into segments and the segments will be sent separately. Each of the segments is attached with a header which is duplicated from the original PDU header. Fields that distinguish between a segment and another are data unit identifier and segment offset. As specified by the ISO 8473-1 [4], the length of the data part of each segment should be a multiplication of 8 octets, except for the last segment.

After a PDU has been properly composed, it is ready to be transmitted to the outgoing network device. Firstly, the physical address of the outgoing and receiving network devices has to be determined and these addresses are attached to the packet as Ethernet header to form Ethernet frame. Then the interface function that invokes the device driver to perform the frame transmission is called.

This research has implemented the mentioned functions. The packet composition is implemented in *clnp_queue_xmit* function. The segmentation function is implemented in *clnp_fragment()* function. The transmission to device function is implemented in *clnp_finish_output()* and *dev_queue_xmit()* methods.

### 3.5. CLNP inside Linux Kernel version 2.6.

The data structure used for implementing CLNP is the same as the one currently used by all the network layers inside Linux kernel version 2.6. This structure, called socket buffer (*sk_buff*), is used by all the network layers to store their headers, information about the user data (the payload), and other information needed internally for coordinating their work [1].

CLNP is integrated into the structure, inside the network layer (*union nh*). After integrating CLNP header inside *sk_buff* structure and having a pointer to a *sk_buff* structure called *skb*, we can access CLNP header with two ways; *skb* → *nh.clnph* and *skb* → *nh.raw[0]*.

## 4. Implementation and Unit Testing

Each input, routing and output process is implemented as separate modules. Tests are done by inserting the functions inside the kernel module, under /usr/src/Linux-2.6.17.13/net/clnp. The clnp header is put under /usr/src/Linux-2.6.17.13/include/linux. A pointer to the clnp header is inserted into the *sk_buff* structure, *union nh*. Then, each module is tested independently. The following are some of the test cases.

**4.1 CLNP Input process**

1. Input test
   a. A PDU comprises of fixed part and address part is received. A PDU comprises of fixed part, address part, segmentation part, and options part is received.
   b. An error report PDU is received.
   c. An inactive network layer PDU is received.
   d. A PDU with a bit error is received.
   e. A PDU which belongs to other end system is received.
2. Error detection and reporting test
   a. A PDU with checksum parameters are zero is received.
   b. A PDU without an error is received.
   c. A PDU with a bit error and ER flag set is received.
3. Reassembly test
   a. Eight out of order segments (without overlap) constructing three initial PDUs are received.
   b. Eight out of order segments (with overlap) constructing three initial PDUs are received.
   c. Seven from eight out of order segments constructing three initial PDUs are received (reassembly timer is expired)

**4.2 CLNP Routing Process**

We simulate a path with 2 end systems (source and destination) and 2 intermediate systems. These are some of the test cases conducted.

1. Source routing function test
   a. Complete and Partial Source Routing with destination address, intermediate system address 1 and 2 are exists in routing table.

b. Complete and Partial Source Routing with intermediate system address 1 and 2 are exists in routing table. But the destination address is not exists in routing table.
   c. Complete Source Routing with destination address exists in routing table but intermediate system address1 or 2 are not exists in routing table. In this test, the condition will be established with one IS address as false address.
   d. Partial Source Routing with destination address and intermediate system address 1 are exists in routing table. But the intermediate system address 2 is not exists in routing table.
   e. Partial Source Routing with intermediate system address 1 exists in routing table. But the destination address and intermediate system address 2 are not exists in routing table.
   f. Partial Source Routing with destination address and intermediate system address 2 are exists in routing table. But the intermediate system address 1 is not exists in routing table.
   g. Partial Source Routing with destination address exists in routing table. But the intermediate system address 1 and 2 are not exists in routing table.

2. Forward PDU function test
   a. Forwarding function in proper condition
   d. Forwarding function if header error occurs
   e. Forwarding function if the packet type is not same with the packet host
   f. Forwarding function if TTL (Time to Live) expires
   g. Forwarding function if the destination is not same with the gateway

## 4.3 CLNP Output Process

1. PDU composition test
   a. Fixed part and address part compostion.
   b. Fixed part, address part, and segmentation part compostion.
   c. Fixed part, address part, segmentation part, and options part compostion.

2. PDU segmentation test
   a. Segmenting a 100-byte PDU.
   b. Segmenting a 200-byte PDU.

3. PDU transmission test
   a. Transmission using dev_queue_xmit function.
   b. Transmission using raw socket.

A makefile for each module is created and compiled, creating modules. Then, each of the modules resulted from the compilation is loaded manually into the Linux kernel using the 'insmod' command. The results for each testing can be seen inside /var/log/messages/.

## 5. Conclusion

The designs of the CLNP input, routing, and output processes have been designed based on ISO 8473-1 and successfully implemented as loadable kernel modules inside Linux version 2.6. Some tests for each module have been conducted and they worked. For further work, the performance of these functions should be measured. Moreover, all modules inside CLNP (input, routing, and output modules) should be integrated and implemented as a complete CLNP module.

# 6. References

[1] Benvenuti, Christian: "Understanding Linux Network Internals", O'Reilly, USA, 2006.
[2] Goutelle, M., Hughes-Jones, R., Kelly, T., Li, Y-T., Martin-Flatin, J. P., Rio, M.: "A Map of the Networking Code in Linux Kernel 2.4.20", 2004.
[3] ICAO: "ASIA/PACIFIC Regional Plan for the New CNS/ATM System", 2005.
[4] ISO/IEC: "Information Technology – Protocol for providing the connectionless-mode network service: Protocol specification", ISO/IEC 8473-1, second edition, ISO/IEC Copyright Office, Switzerland, 1st November 1998.
[5] NetBSD.: "NetBSD CVS Repositories". Accessed on 5th January 2007. Available at http://cvsweb.netbsd.org/bsdweb.cgi/
[6] The Agency for The Assesment and Application of Technology (BPPT): Technology Roadmap of The Agency for BPPT, year 2007-2015. Jakarta, March 2007.